\documentclass[aps,showpacs,showkeys,preprintnumbers,amsmath,amssymb,nofootinbib]{revtex4}
\usepackage{amsmath}
\usepackage{amssymb}
\usepackage{graphics}
\usepackage{eurosym}
\usepackage{amsfonts}
\usepackage{mdframed}
\usepackage{graphicx}
\usepackage[font={footnotesize,it}]{caption}

\begin{document}
\baselineskip=0.6 cm
\title{The Effect of the Gauss-Bonnet term to Hawking Radiation from arbitrary dimensional  Black Brane}
\author{Xiao-Mei Kuang}
\email{xmeikuang@gmail.com}

\affiliation{Instituto de F\'{\i}sica, Pontificia Universidad Cat\'olica de
Valpara\'{\i}so, Casilla 4950, Valpara\'{\i}so, Chile}
\affiliation{Center for Gravitation and Cosmology, College of Physical Science and Technology, Yangzhou University, Yangzhou 225009, China}

\author{Joel Saavedra}
\email{joel.saavedra@pucv.cl}

\affiliation{Instituto de F\'{\i}sica, Pontificia Universidad Cat\'olica de
Valpara\'{\i}so, Casilla 4950, Valpara\'{\i}so, Chile}

\author{Ali \"{O}vg\"{u}n}
\email{ali.ovgun@pucv.cl}

\affiliation{Instituto de F\'{\i}sica, Pontificia Universidad Cat\'olica de
Valpara\'{\i}so, Casilla 4950, Valpara\'{\i}so, Chile}

\affiliation{Physics Department, Eastern Mediterranean University, Famagusta,
Northern Cyprus}

\date{\today }

\vspace*{0.2cm}
\begin{abstract}
\baselineskip=0.6 cm
\begin{center}
{\bf Abstract}
\end{center}
We investigate the probabilities of the tunneling and the radiation spectrums of the massive spin-1 particles from arbitrary dimensional 
Gauss-Bonnet-Axions (GBA) Anti-de Sitter (AdS) black brane, via using the WKB approximation to the Proca spin-1 field equation. The tunneling probabilities and  Hawking temperature of the arbitrary dimensional GBA AdS black brane is calculated via the Hamilton-Jacobi approach. We also compute the Hawking temperature via Parikh-Wilczek tunneling approach. The results obtained from the two methods are consistent. In our setup, the Gauss-Bonnet (GB) coupling affects the Hawking temperature if and only if the momentum of axion fields is non-vanishing.
\end{abstract}

\maketitle
\newpage

\section{Introduction}\label{sec:Introduction}

Since Stephen Hawking proposed that black holes are not actually black, inasmuch as they emit a nearly thermal radiation, and this radiation causes them to lose energy, shrink and eventually disappear. This is the disaster for the laws of quantum physics because information is lost, which is the unsolved black hole information problem \cite{SH1,SH2}. Recently, Hawking, Perry and Strominger show some ways to solve the information problem by defining new term in physics `soft particles' \cite{SH3,SH4} and `hard particles' \cite{strominger}. Soft Particles carry information and imprint this information on the radiation, even after the black hole itself is defunct. 

In the process of constructing quantum gravity theory, physicists keep strong interests in the study of Hawking radiation proposed in \cite{SH1,SH2}. In order to strengthen this staggering theory, in the last decades, plenty of methods have been proposed to derive Hawking radiation and calculate its temperature \cite{Unruh,DR,pw,parik,sri,s1,Clement}. Among them, the semi-classical quantum tunneling method proposed in \cite{pw} has been attracted many attentions and gotten remarkable progress. The trick in the method is to consider the Hawking radiation as a tunneling process of particles from the event horizon, and the tunneling probability for the classically forbidden trajectory to outside of the horizon is $\Gamma=e^{-\frac{2}{\hbar} \mathrm{Im} S}$. So the crucial task is to compute the imaginary part of the classical action $\mathrm{Im} S$.

In the tunneling method, there are two popular approaches to calculate $\mathrm{Im} S$. One is Parikh-Wilczek approach which applies the null geodesic equation of the emitted scalar particles \cite{pw}. This approach has been extensively used in \cite{zhao1,zhao2,zhao3}. The other way is Hamilton-Jacobi approach which mainly utilizes the  Hamilton-Jacobi equation of the classical scalar particles proposed in \cite{ang1,ang2,mann0}. Later, tunneling of Dirac fermions \cite{Mann1,Mann2,Liran1,Liran2,Liran3,Chen1,Jiang} and tunneling of gravitinos \cite{Mann3,Chen2} have been imposed to study the Hawking radiation. More recently, tunneling of massive spin-1 vector particles has been widely studied to probe the radiation, which was first researched in \cite{Kruglov1,Kruglov2}. This kind of tunneling was studied via WKB approximation of the Proca equation which describes the dynamic of spin-1 vector particles. Tunneling of the different types of particles have been generalized to more objects, see for examples \cite{Chen3,Sakalli1,Li,lii,Feng,limo,Sak1,joel1,sing1,sing2,sing3,zeng1,zeng2,masji,Sak0,Sak2,ao2,Sak3,SM0,ao6,SM1,SM3,SM4,VHR4,VHR5,ao3,ao4,ao7,ao5,kj1,kj2,kj3,PAD,corda,biswas,abbas,mhali,feng1,mhali2,mhali3,mhali4,siah,mart,epjc1,gardo,gacim1,gacim2,ahmed1}.

In this paper, we will explore the quantum tunneling process of massive spin-1 vector particles emitted from arbitrary dimensional AdS black branes with GB corrections as well as axion fields founded in\cite{Kuang:2017cgt}\footnote{We note that the  Hawking temperature is impeccably might obtained using tunnelling of scalar, vector or Dirac particles, and the result is independent of the spin type of the particles. In this paper, we will only consider the massive vector particles which should give correct Hawking temperature, because the studying of vector particles is more new and physically reasonable.}. To calculate the tunneling probability and the Hawking temperature, we first use the Hamilton-Jacobi approach by starting with Proca equation, then we verify our results via  Parikh-Wilczek approach.

The motivations of our study stem from the following three aspects. Firstly, the study of tunneling of spin-1 particles is very significant in the particle physics because a vector boson is with spin-1 and the massive vector bosons  $W^{\pm}$ and $Z^0$ particles (force carriers of the weak interaction) play an indispensable role in the confirmed Higgs Boson \cite{RevPar}. Secondly, it is well known that the gravitational theories with higher curvature coupling, such as GB  corrections and the higher dimensions are very important and contain much richer physics. Thirdly, from the point view  of the application of Anti de-Sitter/conformal field theory,  massless axion fields breaks the translation symmetry of the dual boundary theory in a very simple way which was first discussed in \cite{Andrade:2013gsa}, so that it introduces finite DC conductivity in the study of holographic applications.  

As we know that the temperature of GB black branes with panel spatial topology does not depend on the GB parameters\cite{Boulwar,Cai,Cvetic}. However, in the GBA gravitational theory constructed in \cite{Kuang:2017cgt}, via Euclidean method, the authors found  that the temperature of GB AdS panel black hole with axion fields is dependent of the GB coupling. When there is no Axion field, the temperature recovers that in normal GB theory. Thus, the aim of this paper is to calculate the Hawking temperature via tunneling method and further confirm the result obtained in\cite{Kuang:2017cgt}.

The remaining of this paper is organized as follows. In section \ref{sec-background}, we will review the black brane solution in Gauss-Bonnet-Axions theory with arbitrary dimensions. Using the Hamilton-Jacobi approach, we compute the tunneling probabilities of the spin-1 vector  particles and Hawking temperature of the black brane in section \ref{section:approach 1}, and we verify the Hawking  temperature with the use of Parikh-Wilczek approach in section \ref{section:approach 1}. The last section is our conclusion and discussion. 

\section{Review of Black Brane in Gauss-Bonnet-Axions theory with arbitrary dimensions}\label{sec-background} 
In this section, we will review the black
hole solutions in arbitrary dimensional Gauss-Bonnet-Axions(GBA) theory
proposed in \cite{Kuang:2017cgt} with the action 
\begin{equation}
S=\frac{1}{2\kappa^{2}}\int d^{d+2}x\sqrt{-g}\Big(R-2\Lambda+\frac{\alpha}{2}\mathcal{L}_{GB}-\frac{1}{2}\sum_{I=1}^{d}(\partial{\psi_{I}})^{2}\Big).\label{action}
\end{equation}
Here $2\kappa^{2}=16\pi G_{d+2}$ is the $d+2$ dimensional gravitational
coupling constant and $\alpha$ is the GB coupling constant. $\psi_{I}$
are a set of axionic fields and $\Lambda=-d(d+1)/2L^{2}$ is the
cosmological constant while 
\begin{equation}
\mathcal{L}_{GB}=\left(R_{\mu\nu\rho\sigma}R^{\mu\nu\rho\sigma}-4R_{\mu\nu}R^{\mu\nu}+R^{2}\right).
\end{equation}
In what follows, we shall set $L=1$.

Variating the action, we can obtain the equations of motion as 
\begin{eqnarray}
 &  & \nabla_{\mu}\nabla^{\mu}\psi_{I}=0,\nonumber \\
 &  & R_{\mu\nu}-\frac{1}{2}g_{\mu\nu}\Big(R+d(d+1)+\frac{\alpha}{2}(R^{2}-4R_{\rho\sigma}R^{\rho\sigma}+R_{\lambda\rho\sigma\tau}R^{\lambda\rho\sigma\tau})\Big)\nonumber \\
 &  & +\frac{\alpha}{2}\left(2RR_{\mu\nu}-4R_{\mu\rho}R_{\nu}^{~\rho}-4R_{\mu\rho\nu\sigma}R^{\rho\sigma}+2R_{\mu\rho\sigma\lambda}R_{\nu}^{~\rho\sigma\lambda}\right)-\sum_{I=1}^{d}\left(\frac{1}{2}\partial_{\mu}{\psi_{I}}\partial_{\nu}{\psi_{I}}-\frac{g_{\mu\nu}}{4}(\partial{\psi_{I}})^{2}\right)=0.
\end{eqnarray}
We take the form of the scalar fields linearly depending on the $d$
spatial direction $x^{a}$ where $a=1,2,\cdots d$ as 
\begin{equation}
\psi_{I}=\beta\delta_{Ia}x^{a},
\end{equation}
which breaks the translational symmetry in the dual field theory in
a simple way\cite{Andrade:2013gsa}. The equations of motions admit
a homogeneous and isotropic neutral black brane solution with panel topology\footnote{In order to have a unit velocity of light, 
the metric ansatz $g_{aa}$  are dependent of GB coupling parameter $\hat{\alpha}$. They are somehow different from
one presented  in \cite{Cheng:2014tya} in five dimension where $g_{aa}=r^{2}/L^{2}$
is independent of GB coupling $\hat{\alpha}$. }
\begin{equation}
ds^{2}=-f(r)dt^{2}+\frac{1}{f(r)}dr^{2}+\frac{r^{2}}{L_{e}^{2}}\sum_{a=1}^d dx^{a}dx^{a},\label{eq-metric}
\end{equation}
where, with defining $\hat{\alpha}=(d-1)(d-2)\alpha/2$, 
\begin{eqnarray}
f(r)=\frac{r^{2}}{2\hat{\alpha}}\left(1-\sqrt{1-4\hat{\alpha}\left(1-\frac{r_{h}^{d+1}}{r^{d+1}}\right)+\frac{2\hat{\alpha}}{r^{2}}\frac{L_{e}^{2}\beta^{2}}{d-1}\left(1-\frac{r_{h}^{d-1}}{r^{d-1}}\right)}\right).\label{eq-fr}
\end{eqnarray}
Here $r_{h}$ satisfying $f(r_{h})=0$ is the black brane horizon.

The constraint of GB coupling parameter $\hat{\alpha}$ has been widely studied in arbitrary dimensions. Considering holography,  
the physical range of $\hat{\alpha}$ can mainly determined by two constraints. One constraint is to require no negative
energy fluxes appear in all tensor, vector and scalar channels of perturbations, which can be achieved by computing the energy flux via holography and then translate the constraints of no negative energy fluxes condition into the constraint on the GB parameter\cite{Buchel:2009sk}. The other constraint is to demand that the dual conformal field theory of GB theory is causal. It means that the local ``speed of graviton" should not be over than the speed of light, which can be realized via analyzing the potentials of the channels of gravitational perturbations\cite{Brigante:2008gz}. From \cite{Buchel:2009sk,Brigante:2008gz,Camanho:2009vw,Ge:2009eh}, the above two constraints produced the similar physical range of GB coupling parameter, which is 
\begin{equation}
-\frac{(d-1)(3d+5)}{4(d+3)^{2}}\leqslant\hat{\alpha}\leqslant\frac{(d-1)(d-2)((d-1)^{2}+3d+5)}{4((d-1)^{2}+d+3)^{2}}.\label{constraint}
\end{equation}
It is worthy to note that it was shown in \cite{Camanho:2014apa} that the constraint in the
higher derivative coupling coming from causality issues is much more
severe, especially in a weakly coupled theory. And later in \cite{DAppollonio:2015fly},
the authors pointed out that the causality violations can be cured by considering the Regge behavior, and more recent studies about the (in)stability 
can be seen in\cite{Konoplya:2017ymp,Konoplya:2017lhs,Konoplya:2017zwo}. The preciser causality constraint
of higher correction coupling is worthy further investigated. Moreover, the result is given in the case without the axion fields.
It may become more complicate due to the introduction of axionic fields \cite{Wang:2016vmm}.

Near the UV boundary $r\rightarrow\infty$, 
\begin{equation}
f(r)\sim\frac{1-\sqrt{1-4\hat{\alpha}}}{2\hat{\alpha}}r^{2}.
\end{equation}
So the effective asymptotic AdS radius is 
\begin{eqnarray}
L_{{\rm e}}^{2}=\frac{2\hat{\alpha}}{1-\sqrt{1-4\hat{\alpha}}}\to\left\{ \begin{array}{rl}
1\ , & \text{for}\ \hat{\alpha}\rightarrow0\\
\frac{1}{2}\ , & \text{ for}\ \hat{\alpha}\rightarrow\frac{1}{4}
\end{array}\right.\,.
\end{eqnarray}
Then, using the Euclidean method, the Hawking temperature of the black brane is 
\begin{equation}\label{hawking1}
T=\frac{f'(r_{h})}{4\pi}=\frac{1}{4\pi}\left((d+1)r_{h}-\frac{L_{e}^{2}\beta^{2}}{2r_{h}}\right).
\end{equation}
depending on the GB coupling via $L_{e}$ when $\beta \neq 0$, which is very different from the GB gravitational with $\beta=0$.

The Einstein limit of the black brane solution is obtained by taking $\hat{\alpha}\rightarrow0$,
in which the gravitational background recovers the solution addressed
in \cite{Andrade:2013gsa} with the redshift 
\begin{equation}
f(r)=r^{2}\left(1-\frac{r_{h}^{d+1}}{r^{d+1}}\right)-\frac{\beta^{2}}{2(d-1)}\left(1-\frac{r_{h}^{d-1}}{r^{d-1}}\right),
\end{equation}
and the temperature 
\begin{equation}
T=\frac{f'(r_{h})}{4\pi}=\frac{1}{4\pi}\left((d+1)r_{h}-\frac{\beta^{2}}{2r_{h}}\right).\label{hawking2}
\end{equation}

It is worth to point out that dimension of the solution \eqref{eq-fr} can be $d\geqslant 2$ for Einstein case with $\hat{\alpha}\rightarrow0$, while for GB case with $\hat{\alpha}\neq 0$, we have $d\geqslant 3$ since  the Gauss-Bonnet term is topological invariant in four dimension \cite{Cai}.

\section{Quantum Tunneling of Spin-1 Particles from $d+2$ dimensional Black Brane in Gauss-Bonnet-Axions
Theory}\label{section:approach 1}
Our main aim of this section is to calculate the Hawking temperature
using the semi-classical Hamilton-Jacobi approach with the suitable WKB
ansatz from the recently founded solution of the black brane in GBA
theory. In order to calculate the Hawking temperature we consider the quantum
mechanically tunneling massive spin-1 particles from the black brane
in the GBA theory. The wave of the massive spin-1 field $\psi_{\nu}$  satisfies the dynamical Proca equation \cite{Kruglov1}
\begin{equation}
\frac{1}{\sqrt{-g}}\partial_{\mu}\left(\sqrt{-g}\psi^{\nu\mu}\right)+\frac{m^{2}}{\hbar^{2}}\psi^{\nu}=0,\label{10proca}
\end{equation}
where $m$ is the mass of the spin-1 particles and $\psi_{\mu\nu}$
is the $2^{nd}$-rank tensor which is defined by $\psi_{\mu\nu}=\partial_{\mu}\psi_{\nu}-\partial_{\nu}\psi_{\mu}$. Note that here $\mu$ and $\nu$ go through the coordinates as $t,r,x^1,x^2\cdots x^d$.

We expand the Proca equation in the background of the black brane \eqref{eq-metric} in the GBA theory and find that  in arbitrary dimension $D=d+2$, we have the components of the equations
\begin{eqnarray}
&&\nu=t:~~\frac{m^2}{\hbar^2f}\psi_{t}+\frac{d}{r}\left(\frac{\partial\psi_t}{\partial r}-\frac{\partial\psi_r}{\partial t}\right)+\frac{\partial^2\psi_t}{\partial r^2}-\frac{\partial^2\psi_r}{\partial t\partial r}+\frac{L_e^2}{r^2 f}\left(\sum_{a=1}^d\frac{\partial^2\psi_t}{\partial x^{a2}}-\sum_{a=1}^d\frac{\partial^2\psi_{x^a}}{\partial t\partial x^a}\right)=0,\nonumber\\
&&\nu=r:~~\frac{m^2f}{\hbar^2}\psi_{r}+\frac{\partial^2\psi_t}{\partial t\partial r}-\frac{\partial^2\psi_r}{\partial t^2}+\frac{L_e^2 f}{r^2}\left(\sum_{a=1}^d\frac{\partial^2\psi_r}{\partial x^{a2}}-\sum_{a=1}^d\frac{\partial^2\psi_{x^a}}{\partial r\partial x^a}\right)=0,\nonumber\\
&&\nu=x^1:~~\frac{m^2 L_e^2}{\hbar^2r^2}\psi_{x^1}+\frac{L_e^4}{r^4}\left(\sum_{a=2}^d\frac{\partial^2\psi_{x^1}}{\partial x^{a2}}-\sum_{a=2}^d\frac{\partial^2\psi_{x^a}}{\partial x^1\partial x^a}\right)\nonumber\\
&&~~~+\frac{L_e^2}{r^2}\left[\left(\frac{\partial \psi_{x^1}}{\partial r}-\frac{\partial \psi_{r}}{\partial x^1}\right)\left(f'+\frac{(d-2)f}{r}\right)+f\left(\frac{\partial \psi^2_{x^1}}{\partial r^2}-\frac{\partial \psi^2_{r}}{\partial r\partial x^1}\right)+\frac{1}{f}\left(\frac{\partial^2 \psi_{t}}{\partial t\partial x^1}-\frac{\partial^2 \psi_{x^1}}{\partial t^2}\right)\right]=0,\nonumber\\
&&~~~~~~~~~~~~~~~~~~~~~~~~~~~~~~~~~~~~~~~~~~~~~~~~~~\vdots\nonumber\\
&&\nu=x^j:~~\frac{m^2 L_e^2}{\hbar^2r^2}\psi_{x^j}+\frac{L_e^4}{r^4}\left(\sum_{a=1,a\neq j}^d\frac{\partial^2\psi_{x^j}}{\partial x^{a2}}-\sum_{a=1,a\neq j}^d\frac{\partial^2\psi_{x^a}}{\partial x^j\partial x^a}\right)\nonumber\\
&&~~~+\frac{L_e^2}{r^2}\left[\left(\frac{\partial \psi_{x^j}}{\partial r}-\frac{\partial \psi_{r}}{\partial x^j}\right)\left(f'+\frac{(d-2)f}{r}\right)+f\left(\frac{\partial \psi^2_{x^j}}{\partial r^2}-\frac{\partial \psi^2_{r}}{\partial r\partial x^j}\right)+\frac{1}{f}\left(\frac{\partial^2 \psi_{t}}{\partial t\partial x^j}-\frac{\partial^2 \psi_{x^j}}{\partial t^2}\right)\right]=0,\label{eq:soln}
\end{eqnarray}
where the index $j=2,3\cdots d$ and the prime denotes the derivative to the radial coordinate $r$.
Then using the Hamilton-Jacobi method, we choose the suitable ansatz of the spin-1 vector
field \cite{Kruglov2}
\begin{equation}
\psi_{\nu}=C_{\nu}\exp\left[\frac{i}{\hbar}\left(S_{0}(t,r,x^1,x^2,\cdots,x^d)+\hbar\,S_{1}(t,r,x^1,x^2,\cdots,x^d)+\cdots.\right)\right],\label{12an}
\end{equation}
where $C_{\nu}=(C_{t},C_{r},C_{x^1},C_{x^2},\cdots,C_{x^d})$ are some constants. It is noted that a classical action of the massive spin-1 particles are defined by $S_{0}(t,r,x^1,x^2,\cdots,x^d)$ which is a kinetic term \cite{Vanzo}.
Higher order correction terms  $S_{i=1,2,..}(t,r,x^1,x^2,\cdots,x^d)$ can be ignored because the particles can be considered to be free and without self-interacting. Moreover, to choose the suitable WKB ansatz, one should
use the symmetry of the background spacetime by Killing vectors \cite{mann0}.
In our case, we choose the action in leading orders as follows
\begin{equation}
S_{0}(t,r,x^1,x^2,\cdots)=-Et+R(r)+\sum_{a=1}^d X^a(x^a)+c,\label{13act}
\end{equation}
which means that the particles of massive spin-1 have a energy of $E$. Here $c$ is a complex constant. Then, we substitute Eq. \eqref{12an} and Eq. \eqref{13act} into the components of  Proca equation given in Eq. \eqref{eq:soln}, the corresponding quadruple equations with the lowest order in $\hbar$
can be obtained 
\begin{subequations}\label{lower-hbar}
\begin{eqnarray}\label{lower-hbarA}
&&\left[ \frac{m^2}{f}+\left(\frac{\partial R}{\partial r}\right)^2-\frac{L_e^2}{r^2 f}\sum_{a=1}^d\left(\frac{\partial X^a}{\partial x^a}\right)^2  \right]C_t-E\frac{\partial R}{\partial  r}C_r-\frac{L_e^2E}{r^2 f} \sum_{a=1}^d\left( \frac{\partial X^a}{\partial x^a}C_{x^a}\right)=0,
\end{eqnarray}
\begin{eqnarray}\label{lower-hbarB}
&&E\frac{\partial R}{\partial r} C_t+\left[m^2 f+E^2+\frac{L_e^2 f}{r^2}\sum_{a=1}^d\left( \frac{\partial X^a}{\partial x^a}\right)^2 \right]C_r+\frac{L_e^2 f}{r^2}\frac{\partial R}{\partial r}\sum_{a=1}^d\left(\frac{\partial X^a}{\partial x^a}C_{x^a}\right)=0,
\end{eqnarray}
\begin{eqnarray}\label{lower-hbarC}
&&\frac{L_e^2 E}{r^2 f}\frac{\partial X^1}{\partial x^1}C_t+\frac{L_e^2 f}{r^2 }
\frac{\partial R}{\partial r}\frac{\partial X^1}{\partial x^1}C_r+\frac{L_e^4 }{r^4}\frac{\partial X^1}{\partial x^1}\sum_{a=2}^d\left(\frac{\partial X^a}{\partial x^a}C_{x^a}\right)\nonumber\\
&&+\left[  \frac{m^2 L_e^2}{ r^2}-\frac{L_e^4}{r^4}  \sum_{a=2}^d\left( \frac{\partial X^a}{\partial x^a}\right)^2 + \frac{L_e^2}{r^2}\left(\frac{E^2}{f}-f\left(\frac{\partial R}{\partial r}\right)^2 \right)  \right]C_{x^1}=0,
\end{eqnarray}
~~~~~~~~~~~~~~~~~~~~~~~~~~~~~~~~~~~~~~~~~~~~~~~~~~\vdots
\begin{eqnarray}\label{lower-hbarD}
&&\frac{L_e^2 E}{r^2 f}\frac{\partial X^j}{\partial x^j}C_t+\frac{L_e^2 f}{r^2 }
\frac{\partial R}{\partial r}\frac{\partial X^j}{\partial x^j}C_r+\frac{L_e^4 }{r^4}\frac{\partial X^j}{\partial x^j}\sum_{a=1, a\neq j}^d\left(\frac{\partial X^a}{\partial x^a}C_{x^a}\right)
\nonumber\\
&&+\left[  \frac{m^2 L_e^2}{ r^2}-\frac{L_e^4}{r^4}  \sum_{a=1, a\neq j}^d\left( \frac{\partial X^a}{\partial x^a}\right)^2 + \frac{L_e^2}{r^2}\left(\frac{E^2}{f}-f\left(\frac{\partial R}{\partial r}\right)^2 \right)  \right]C_{x^j}=0
\end{eqnarray}
\end{subequations}
where again $j=2,3\cdots d$.

We will use the matrix formalism to solve these $d+2$ equations for
radial function of $R(r)$. Let's first rewrite Eq.\eqref{lower-hbar}
in a $(d+2)\times (d+2)$ matrix form as $\varLambda_{(d+2)\times (d+2)}\left(C_{t},C_{r},C_{x^1}\cdots C_{x^d}\right)^{\mathcal{T}}=0$,
where $\mathcal{T}$ is the transition to the transposed vector. It is notable that here $\varLambda_{(d+2)\times (d+2)}$ is in matrix form and the function of the coordinates, and indeed $C_{\mu}$ are constants defined in equation \eqref{12an}.  So we can read off all the elements of the matrix $\varLambda_{(d+2)\times (d+2)}$ from the related \eqref{lower-hbarA}-\eqref{lower-hbarD} as
\begin{subequations}
\begin{eqnarray}
\varLambda_{1,1}&=&  \frac{m^2}{f}+\left(\frac{\partial R}{\partial r}\right)^2-\frac{L_e^2}{r^2 f}\sum_{a=1}^d\left(\frac{\partial X^a}{\partial x^a}\right)^2 ,~~~~ \varLambda_{1,2}=-E\frac{\partial R}{\partial  r}
\nonumber\\ \varLambda_{1,3}&=&-\frac{L_e^2E}{r^2 f} \frac{\partial X^1}{\partial x^1},~~~\cdots,~~~
 \varLambda_{1,d+2}=-\frac{L_e^2E}{r^2 f} \frac{\partial X^d}{\partial x^d}.
\end{eqnarray}
\begin{eqnarray}
\varLambda_{2,1}&=&E\frac{\partial R}{\partial r},~~~~ \varLambda_{2,2}=m^2 f+E^2+\frac{L_e^2 f}{r^2}\sum_{a=1}^d\left(\frac{\partial X^a}{\partial x^a}\right)^2   
\nonumber\\ \varLambda_{2,3}&=&\frac{L_e^2 f}{r^2}\frac{\partial R}{\partial r}\frac{\partial X^1}{\partial x^1},~~~\cdots,~~~
 \varLambda_{2,d+2}=\frac{L_e^2 f}{r^2}\frac{\partial R}{\partial r} \frac{\partial X^d}{\partial x^d}.
\end{eqnarray}
\begin{eqnarray}
\varLambda_{3,1}&=&\frac{L_e^2 E}{r^2 f}\frac{\partial X^1}{\partial x^1},~~~~ \varLambda_{3,2}=\frac{L_e^2 f}{r^2 }\frac{\partial R}{\partial r}\frac{\partial X^1}{\partial x^1}
\nonumber\\ \varLambda_{3,3}&=&   \frac{m^2 L_e^2}{ r^2}-\frac{L_e^4}{r^4}  \sum_{a=2}^d\left( \frac{\partial X^a}{\partial x^a}\right)^2 + \frac{L_e^2}{r^2}\left(\frac{E^2}{f}-f\left(\frac{\partial R}{\partial r}\right)^2 \right)    \nonumber\\
\nonumber\\ \varLambda_{3,4}&=& \frac{L_e^4}{r^4}\frac{\partial X^1}{\partial x^1} \frac{\partial X^2}{\partial x^2},
  ~~~\cdots,~~~
 \varLambda_{3,d+2}=\frac{L_e^4}{r^4}\frac{\partial X^1}{\partial x^1} \frac{\partial X^d}{\partial x^d}.
\end{eqnarray}
~~~~~~~~~~~~~~~~~~~~~~~~~~~~~~~~~~~~~~~~~~~~~~~\vdots
\begin{eqnarray}
\varLambda_{j+2,1}&=&\frac{L_e^2 E}{r^2 f}\frac{\partial X^j}{\partial x^j},~~~~\varLambda_{j+2,2}=\frac{L_e^2 f}{r^2 }\frac{\partial R}{\partial r}\frac{\partial X^j}{\partial x^j}
\nonumber\\ \varLambda_{j+2,3}&=& \frac{L_e^4}{r^4}\frac{\partial X^j}{\partial x^j} \frac{\partial X^1}{\partial x^1} ,~~~\cdots~~~
 \varLambda_{j+2,j+1}= \frac{L_e^4}{r^4}\frac{\partial X^j}{\partial x^j} \frac{\partial X^{j-1}}{\partial x^{j-1}}   \nonumber\\
\nonumber\\ \varLambda_{j+2,j+2}&=&   \frac{m^2 L_e^2}{ r^2}-\frac{L_e^4}{r^4}  \sum_{a=1,a\neq j}^d\left( \frac{\partial X^a}{\partial x^a}\right)^2 + \frac{L_e^2}{r^2}\left(\frac{E^2}{f}-f\left(\frac{\partial R}{\partial r}\right)^2 \right)     \nonumber\\
\nonumber\\ \varLambda_{j+2,j+3}&=& \frac{L_e^4}{r^4}\frac{\partial X^j}{\partial x^j} \frac{\partial X^{j+1}}{\partial x^{j+1}} ,~~~\cdots,~~~
 \varLambda_{j+2,d+2}=\frac{L_e^4}{r^4}\frac{\partial X^j}{\partial x^j} \frac{\partial X^d}{\partial x^d}.
\end{eqnarray}\\
\end{subequations}
The condition of the matrix equation  have non-trivial solution is $\det\varLambda=0$, which will give us a simple expression
\begin{equation}
L_{e}^{4}\left(-\left(\frac{{\rm d}}{{\rm \partial}r}R\right)^{2}f^{2}r^{2}-L_{e}^{2}f \sum_{a=1}^d\left( \frac{\partial X^a}{\partial x^a}\right)^2+r^{2}\left(-m^{2}f+E^{2}\right)\right)^{3}m^{2}=0.
\end{equation}
It is obvious that the solution to the above equation for the radial function is 
\begin{equation}
R_{\pm}=\int dr\left(\pm\frac{1}{f}\sqrt{E^{2}-\frac{L_{e}^{2}f}{r^{2}}\sum_{a=1}^d\left( \frac{\partial X^a}{\partial x^a}\right)^2-fm^{2}}\right).\label{integ}
\end{equation}
Note that $R_{+}$ is for the outgoing massive spin-1 particles while $R_{-}$ is for the ingoing ones.  It is obvious that there is a pole on the event horizon. To solve this singularity in the integral of the imaginary
part of $R_{\pm}$ in Eq. \eqref{integ}, we use the residue theorem and complex path
integration \cite{ang1,ang2}.
Consequently, the result of the integral is obtained as
\begin{equation}
\mathrm{Im}R_{\pm}=\pm\left.\frac{\pi}{\frac{{\rm d}}{{\rm d}r}f}E\right\vert _{r=r_{h}}+k\label{30}
\end{equation}
 where $k$ is a complex integration constant. Then we calculate the tunneling probabilities of the ingoing and outgoing
massive spin-1 particles 
\begin{align}
P_{outgoing} & =e^{-\frac{2}{\hbar}\mathrm{Im}S_{+}}=e^{-2(\mathrm{Im}R_{+}+\mathrm{Im}k)},\label{32}\\
P_{ingoing} & =e^{-\frac{2}{\hbar}\mathrm{Im}S_{-}}=e^{-2(\mathrm{Im}R_{-}+\mathrm{Im}k)},\label{27}
\end{align}
where we have set $\hbar=1$ in the second equalities.
In agreement with the event horizon of the black brane in the GBA
theory, particles which are ingoing must be absorbed completely, so that
one can choose that $P_{ingoing}=1$. This concern can be achieved
by imposing that $\mathrm{Im}k=-\mathrm{Im}R_{-}$. Further recalling $R_{+}=-R_{-}$ in \eqref{integ},
 the quantum tunneling rate $\Gamma$ of the massive spin-1 particles has the form
\begin{equation}
\Gamma=P_{emission}=\exp\left(-4\mathrm{Im}R_{+}\right)=\exp\left(-\left.\frac{4\pi}{\left(\frac{{\rm d}}{{\rm d}r}f\right)}E\right\vert _{r=r_{h}}\right).\label{28}
\end{equation}
We recall from \cite{Vanzo} that the equivalent of the tunneling rate $\Gamma$ satisfies the Boltzmann equation  $\Gamma=e^{-\beta E}$ with the Boltzmann factor $\beta$ and the Hawking temperature is defined as $T=\frac{1}{\beta}$ . Then we
derive the surface temperature of the black brane 
\begin{equation}\label{hawking}
T=\left.\frac{\left(\frac{{\rm d}}{{\rm d}r}f\right)}{4\pi}\right\vert _{r=r_{h}}=\frac{1}{4\pi}\left((d+1)r_{h}-\frac{L_{e}^{2}\beta^{2}}{2r_{h}}\right).
\end{equation}
This agrees well with the Hawking temperature found in the Eq. (\ref{hawking1}) via Euclidean method.

\section{Hawking temperature via Parikh-Wilczek Tunneling approach}\label{section:approach 2}
In this section, we use the Parikh-Wilczek Tunneling(PWT) approach to find the Hawking temperature of the
black brane solution Eq.\eqref{eq-metric}. In the  PWT approach, one thinks the tunneling particle as a spherical shell which does not have motion in ($\theta ,\varphi $)-directions. That's to say, we can consider the radial null geodesics of a test particle as a massless spherical shell. With massless particle $m=0$, we would presumably say that the particle associated with the wave (photon, or graviton in this case) is massless, and thus they travel along null geodesics. Recently, it is found that there are several phenomena that have the same behaviour as quantum tunnelling, and thus can be accurately described by tunnelling. Examples include the tunnelling of a classical wave-particle association\cite{Eddi}. Moreover, the particle feels itself as a barrier, because when the particle tunnels to outside, the radius of the black hole smaller depending on the the energy of the outgoing particle. Then using the emission and absorption probabilities of the ingoing and the outgoing particles, one can calculate the ratio \cite{pw,parik}
\begin{equation}
\Gamma=\frac{P_{emission}}{P_{absorption}}=e^{-\frac{E}{T_{H}}}=e^{-2\mathrm{Im}S},  \label{12n}
\end{equation}
where  $\mathrm{Im}S$ is for the net imaginary part of action ($\mathrm{Im}S=\mathrm{Im}S_{out}-\mathrm{Im}S_{in}$). Note that another way to get Eq. \eqref{12n} is to use the following connections in the WKB limit, which are written in terms of the imaginary part of the action of the particles, $\mathrm{Im}S_{out}$ and $\mathrm{Im}S_{in}$, as
\begin{equation}
P_{emission}=e^{-2\mathrm{Im}S_{out}}, \ \ \ \ \ \ \ %
P_{absorption}=e^{-2\mathrm{Im}S_{in}}.  \label{13n}
\end{equation}

To proceed, we transform our metric to the non-singular coordinates of Painleve-Gullstrand coordinates (PGCs) with the transformation \cite{pw}
\begin{equation}
d\tau=dt+\frac{\sqrt{1-f(r)}}{f(r)}dr.  \label{9n}
\end{equation}%
It is noted that, herein we use $\tau$ as a new time in the PGCs to measure a proper time.  Then, it is easy to rewrite the metric Eq. \eqref{eq-metric} into the form
\begin{equation}
ds^{2}=-f(r)d\tau^{2}+2\sqrt{1-f(r)}d\tau dr+dr^{2}+\frac{r^{2}}{L_{e}^{2}}dx^{a}dx^{a}.  \label{10}
\end{equation}
In the PGCs, the function of the metric is regular, and the singularity is removed at $r=r_{h}$ (i.e., $f=0$). 
The only radial null geodesics 
\begin{equation}
\dot{r}=\frac{dr}{d\tau}=\pm 1-\sqrt{1-f(r)}. \label{11n}
\end{equation}%
is sufficient to calculate Hawking radiation. The imaginary part of the action for the outgoing waves are given by \cite{limo}\begin{equation}
\mathrm{Im}S_{out}=\mathrm{Im}\int_{r_{in}}^{r_{out}}p_{r}dr
=\mathrm{Im}\int_{r_{in}}^{r_{out}}\int_{0}^{p_{r}}dp_{r}^{\prime }dr,  \label{16n}
\end{equation}%
where $p_{r}$ is the canonical momentum, $%
r_{in}$ and $r_{out}$ are the initial and final radius of the black hole. The particle tunnels through between these radius so the potential barrier is located between these radius.

The total mass of the system ($M$) is fixed inasmuch as the black hole shrinks ($r_{in}>r_{out}$) after Hawking radiation so that black hole can fluctuate. Moreover, in this study, we use only chargeless particles which has a thin-spherical-shell of energy $\omega $. Then we consider the decreasing of the mass of the black hole $M\rightarrow $ $M-\omega $, because of the  effect of the self-gravitation. Using the Hamilton's equation, the momentum is transferred to the energy $%
\dot{r}=\frac{dH}{dp_{r}}$, and the Eq. \eqref{16n} reduces to
\begin{eqnarray}
\mathrm{Im}S_{out}=\mathrm{Im}\int_{r_{in}}^{r_{out}}\int_{M}^{M-\omega }\frac{dr%
}{\dot{r}}dH.  \label{17n}
\end{eqnarray}
One can now use the energy of the particle $\omega $ instead of the $H$ to transform $H=M-\omega ^{\prime }$ to $dH=-d\omega
^{\prime }$  as
\begin{eqnarray}
\mathrm{Im}{S}_{out}=\int_{r_{in}}^{r_{out}}\int_{0}^{\omega }\frac{dr%
}{\dot{r}}(-d\omega ^{\prime })
\ = -\omega \int_{r_{in}}^{r_{out}}\frac{dr}{ 1-\sqrt{1-f(r)}}.
   \label{18n}
\end{eqnarray}
where we used Eq.\eqref{11n} in the second equality.
To solve the above integral, we use the contour integral by ensuring that positive energy solutions decaying in time because there is a pole at the horizon where $\dot r=0$. For the tunneling particle, we obtain the imaginary part of the action
\begin{equation}
\mathrm{Im}S_{out}=\frac{2 \pi \omega}{f'(r_{h})}+O(\omega^2).  \label{21n}
\end{equation}
 On the other hand, the ingoing particle can be ignored, because the probability amplitude of the ingoing particle is unity so we use the above tunneling rate for a outgoing particle from the horizon. Then we find that the tunneling rate is
\begin{equation}
\Gamma =\exp (-4\pi \omega/ f'(r_{h})), \label{24n}
\end{equation}
and following the Hawking temperature is  
\begin{equation} \label{25n}
T=\left.\frac{\left(\frac{{\rm d}}{{\rm d}r}f\right)}{4\pi}\right\vert _{r=r_{h}}=\frac{1}{4\pi}\left((d+1)r_{h}-\frac{L_{e}^{2}\beta^{2}}{2r_{h}}\right).
\end{equation}
This expression of the Hawking temperature is the same as the Eq.\eqref{hawking} and \eqref{hawking1}. Hence, we have correctly recovered the Hawking temperature. 

\section{conclusion}\label{sec:conclusion}
In this paper, we have studied the tunneling of spin-1 particles from the arbitrary dimensional AdS black branes with GB corrections and additional axion fields, using the Hamilton-Jacobi approach to recover the corresponding Hawking temperature. And then we have checked our results using the Parikh-Wilczek approach. Firstly, to investigate the tunneling rate of the spin-1 particles, we have derived the solutions of the Proca equation on the background of the $d+2$ dimensional black brane in Gauss-Bonnet-Axions theory with the help of Hamilton-Jacobi approach. To solve these solutions, we have used the separation of variables and extracted the radial part of the wave solution after taking zero to the determinant of the matrix. Then we have obtained the solution of the complex integral on the event horizon using the residue method. The tunneling rate of the spin-1 particles has been calculated and the Hawking temperature is correctly confirmed compared with the Boltzmann factor. Secondly, we have used the Parikh-Wilczek approach which involves the null geodesic equation of the emitted scalar particles for derivation the Hawking temperature.

 Remarkably, we have showed that the radiation spectra are not purely thermal. There is an effect of the Gauss-Bonnet term on the radiation spectra that means the information can be extracted using this term. This is an important result because it is know that GB parameters have no effect on the temperature of GB black branes with panel topology\cite{Boulwar,Cai,Cvetic}. However, in this model of the GBA gravitational theory \cite{Kuang:2017cgt}, it is clear that the GB coupling term affects the Hawking temperature as well as the information paradox. One can store all information of the black brane inside the GB term to survive it from the lost of information through tunneling from the event horizon. It is noted that if one removes the axions field, the temperature of the normal GB theory is recovered \cite{Kuang:2017cgt}. We can conclude that the amount of the Hawking temperature depends on the values of the parameters of GB. It is worth to mention that we have neglected the back-reaction effects due to the radiation of particles and also self- gravitating effects. We have calculated the Hawking temperature only in a leading term. This work shows that the gravitational theories with higher curvature coupling, such as GB  corrections and the higher dimensions are significantly influential.
  
Lastly, Einstein's theory of gravity concludes that black holes have no hair. Recently, using the datas of the gravitational waves from the LIGO, information of the hair of the black hole is successfully extracted \cite{ligo}. In future, this  kind of experiments may also give light to the extraction information from black holes/branes and solve the information paradox.

\begin{acknowledgments}
This work is partly supported by Chilean FONDECYT grant No.3150006 (X.M. Kuang) and No.3170035 (A. \"{O}vg\"{u}n).
\end{acknowledgments}

\end{document}